\documentclass[a4paper,usenatbib]{mn2e}

\usepackage{hyperref}
\usepackage{amsfonts}
\usepackage{amssymb}
\usepackage{amsmath}
\usepackage{graphicx}
\usepackage[totalwidth=506pt,totalheight=680pt]{geometry}

\newcommand{\adsurl}[1]{\href{#1}{ADS}}
\providecommand{\url}[1]{\href{#1}{#1}}

\newcommand{\avg}[1]{\ensuremath{\langle \,#1\, \rangle}}

\newcommand{\eqn}[1]{equation~\eqref{#1}}

\newcommand{\dd}{\mathrm{d}}
\newcommand{\be}{\begin{equation}}
\newcommand{\ee}{\end{equation}}

\newcommand{\G}{\mathrm{G}}

\newcommand{\MS}{\mathrm{up}}
\renewcommand{\L}{\mathrm{L}}
\newcommand{\DL}{\Delta_\mathrm{L}}

\title[Stochastic excursion sets]
      {Stochasticity in halo formation and the excursion set approach}

\author[M.~Musso, R.~K.~Sheth]{%
Marcello Musso$^{1}$\thanks{E-mail: marcello.musso@uclouvain.be} 
\& Ravi K.~Sheth$^{2,3}$\thanks{E-mail: sheth@ictp.it} \\
 $^1$ CP3-IRMP, Universit\'e Catholique de Louvain, 
      2 Chemin du Cyclotron, 1348 Louvain-la-Neuve, Belgium \\
 $^2$ The Abdus Salam International Center for Theoretical Physics,
      Strada Costiera, 11, Trieste 34151, Italy\\
 $^3$ Center for Particle Cosmology, University of Pennsylvania, 
      209 S. 33rd St., Philadelphia, PA 19104, USA}

\begin{document}

\pagerange{\pageref{firstpage}--\pageref{lastpage}}

\maketitle 

\label{firstpage}

\begin{abstract}
The simplest stochastic halo formation models assume that the traceless part of the shear field acts to increase the initial overdensity (or decrease the underdensity) that a protohalo (or protovoid) must have if it is to form by the present time. Equivalently, it is the difference between the overdensity and (the square root of the) shear that must be larger than a threshold value. To estimate the effect this has on halo abundances using the excursion set approach, we must solve for the first crossing distribution of a barrier of constant height by the random fluctuations of this difference, which is (even for Gaussian initial conditions) a non-Gaussian variate, since the shear is drawn from a $\chi^2_5$ distribution. The correlation properties of such non-Gaussian walks are inherited from those of the density and the shear, and, since they are independent processes, the solution is in fact remarkably simple. We show that this provides an easy way to understand why earlier heuristic arguments about the nature of the solution worked so well. In addition to modelling halos and voids, this potentially simplifies models of the abundance and spatial distribution of filaments and sheets in the cosmic web. 
\end{abstract}

\begin{keywords}
large-scale structure of Universe
\end{keywords}


\section{Introduction}


The abundance and spatial distribution of gravitationally bound objects is a sensitive probe of the nature of the initial conditions, the expansion history of the Universe, and the nature of gravity.  The simplest models of such objects, which we will call halos, assume that they form from the spherically symmetric collapse of sufficiently overdense spherical patches in the primordial fluctuation field \citep{gg72}.  Building on insights from \citet{ps74}, the excursion set approach \citep{bcek91} provides a framework for linking halos to such overdense regions in the primordial field.  In this approach, concentric spheres are assumed to remain concentric as the protohalo collapses around its centre of mass, so one is interested in the largest sphere whose mean overdensity $\delta_\L$ (assumed to be a Gaussian variate) exceeds a critical value $\delta_c$.  

However, halos are not spherical, and in the simplest models of non-spherical collapse, the shear field is assumed to play an important role \citep{bm96,delpopolo98}.  Measurements of halo formation in simulations show that the shear field does indeed matter \citep{smt01}:  it acts to increase the overdensity required for collapse, approximately as 
\begin{equation}
 \label{dcq}
 \delta_\L > \delta_c\, (1 + \sqrt{q^2/q_c^2})
\end{equation}
where $q^2$ is the traceless shear associated with the protohalo patch, and $q_c$ is a parameter that determines how important the effects of the shear are relative to the spherical collapse model.  Large $q_c$ means that the shear must be large if it is to affect halo formation, and spherical collapse is recovered in the $q_c\to\infty$ limit.  Measurements of protohalo patches in simulations suggest that $q_c^2\sim 6\delta_c^2$ \citep{dts13, scs13}.  

The effect of the shear can be incorporated into the excursion set approach by searching for the largest scale on which \eqn{dcq} is satisfied.  The analysis is simplified by the fact that, in a Gaussian random field, $q^2$ is not correlated with $\delta_\L$ \citep{st02}.  However, analytic progress has been hampered by the fact that on each scale $q^2$ is not a Gaussian variate -- if it were, the analysis would be simple \citep[see][]{cs13} -- but is drawn from a $\chi^2_5$ distribution. 

The first crossing problem can be solved numerically of course, by noting that $q^2 = \sum_{i=1}^5 g_i^2/5$ where the $g_i$ are independent Gaussian variates with $\avg{g_i^2} = \avg{\delta_\L^2}$, so the task reduces to generating the six Gaussian walks (one for $\delta_\L$ and the other five to obtain $q$), and checking at each step if \eqn{dcq} is satisfied.  This makes the first crossing problem appear to be six-dimensional, since it depends on six Gaussian walks.  Accounting for the fact that each of these walks has correlated steps is an additional complication.  

The main goal of this work is to show that significant analytic progress can be made by noting that, if one defines $\delta\equiv \delta_\L - q(\delta_c/q_c)$, the multi-dimensional Gaussian problem reduces to that of the single non-Gaussian variate $\delta$ first exceeding $\delta_c$.  One can therefore make use of recent progress in our understanding of the correlated steps problem for non-Gaussian walks \citep{ms12,ms13a}.  In Section~\ref{sec:main} we show that, in fact, this particular problem is even simpler than that for generic non-Gaussian walks, because the walks which make up $\delta$ are themselves Gaussian.  A final section compares our analysis with previous more heuristic approximations, and summarizes.


\section{First crossing distribution with correlated steps}
\label{sec:main}


In the excursion set approach, one is interested in the probability that the average $\delta_\L(r)$ of the overdensity field over a sphere of radius $r$ exceeds the threshold $b$, while for all $R>r$ it remains below $b$. As $r$ changes, $\delta_\L(r)$ describes a random trajectory, whose value at given $r$ has a Gaussian distribution with variance
\begin{equation}
  s(r) \equiv\avg{\delta_\L^2(r)}
  = \int \frac{\dd k}{k}\, \frac{k^3P(k)}{2\pi^2} \, W^2(kr)\,,  
\end{equation}
where $P(k)$ is the power spectrum of $\delta$, and $W(kr)$ is the Fourier transform of the filter that one uses to compute the mean value.
The variance $s$ grows monotonically as $r$ gets smaller, starting from $s=0$ at very large $r$.

In practice, it is convenient to study the walks as a function of $s$ rather than $r$, as this has the advantage of hiding the dependence on the power spectrum and the smoothing filter. One then wants the probability $f(s)$ that 
$\delta(s) > b(s)$ at $s$ but $\delta(S) < b(S)$ for all $S<s$. 
In general, imposing the first constraint is straightforward, whereas the second one is difficult to treat analytically. This difficulty is due to the fact that, for any choice of $W(kr)$ other than a step function in Fourier space, the steps of the walks are correlated with each other.

\subsection{Up-crossing rather than first-crossing}
Since a walk that is first crossing is necessarily reaching the barrier from below, one may begin to approach the problem imposing the less restrictive constraint that $\delta=b$ and the increment $v\equiv \dd\delta/\dd s$ of the walk with scale (the ``velocity'' of the walk) is larger than the increment $b'=\dd b/\dd s$ of the barrier.
This formulation correctly discards walks that are crossing downwards at $s$, although clearly fails to discard those walks that are crossing upwards at $s$ but had already done so at a some larger scale $S$ (i.e.~walks with more than one upcrossing).  
However, \citet{ms12} showed that at small $s$ the fraction of such walks is tiny, since the correlations between steps make sharp turns very unlikely, and walks with more than one upcrossing necessarily take at least two turns. Therefore, the upwards approximation already provides a good approximation to $f(s)$ on the range of scales of interest for Cosmology. Corrections at small $s$, if needed, can be computed as a perturbative expansion in the number of times a walk crosses the barrier going upwards \citep{ms13a} or, non perturbatively, imposing that $f(s)$ is normalized to unity \citep{ms13b}. 

If earlier upcrossings can be neglected, then $f(s)$ can be computed from the joint probability $p(\delta,v;s)$ that a walk reaches $\delta$ at scale $s$ with velocity $v$.  In particular, since one only wants walks that are crossing the barrier upwards, that is $\delta = b(s)$ and $v \ge b'$ (for a barrier of constant height, this is just $v\ge 0$), the first-crossing probability is well approximated by 
\begin{equation}
  f(s) \simeq f_\MS(s)
   \equiv \int_{b'}^{\infty} \!\!\mathrm{d}v\,(v-b')\, p (b,v;s)\,,
\label{fmeanv}
\end{equation}
where the factor of $v-b'$ can be understood as the density current of the upcrossing walks \citep{ms12}. This expression correctly reduces to that of \citet{ps74} in the small-$s$ limit.

Although for a Gaussian distribution evaluating this integral is straightforward, it is in general convenient to use the rescaled stochastic quantities
\begin{equation}
 \Delta\equiv\frac{\delta}{\sqrt{s}},\quad
 \Delta' \equiv \frac{\dd\Delta}{\dd s}\quad  {\rm and}\quad 
 \xi \equiv -\frac{\Delta'}{\sqrt{\avg{\Delta'^2}}}
     \equiv -2\Gamma s\,\Delta'
\end{equation}
where $\Gamma$, defined by $(2\Gamma s)^2\equiv 1/\avg{\!\Delta'^2\!}$, is a weak function of $s$ \citep[e.g.][]{ms12}.  Note that
\begin{equation}
 \avg{\Delta^2}=\avg{\xi^2}=1 \qquad {\rm and}\qquad 
 \avg{\Delta\,\xi}=0\,,
\end{equation}
i.e. $\Delta$ and $\xi$ are uncorrelated (although in a generic non-Gaussian case not independent) random variables. 
Similarly, we will work with
\begin{equation}
  B(s)\equiv \frac{b(s)}{\sqrt{s}}\qquad {\rm and}\qquad
  X \equiv -\frac{\dd B/\dd s}{\sqrt{\avg{\Delta'^2}}}
  = -2\Gamma s\, B',
\end{equation}
where $B' \equiv \dd B/\dd s$.
The sign of $X$ is chosen so that a barrier that does not vary much, as it is typically the case at small $s$, has $X>0$.

\subsection{Gaussian walks}
If $\delta$ is a Gaussian process, then $\Delta$ and $\xi$ are independent random variables and their joint distribution factorizes:  
\begin{equation}
  p(B,\xi) = p(B)\,p(\xi)
              = \frac{{\rm e}^{-B^2/2}}{\sqrt{2\pi}}\,
                \frac{{\rm e}^{-\xi^2/2}}{\sqrt{2\pi}}\,.
\label{pgauss}
\end{equation}
Inserting this in \eqn{fmeanv} shows that $f(s)$ will be 
\begin{equation}
  f_\MS(s) = -B'\,p(B)
  \left[\frac{1 + {\rm erf}(X/\sqrt{2})}{2} 
        + \frac{{\rm e}^{-X^2/2}}{\sqrt{2\pi}X}\right],
\label{sfs}
\end{equation}
where $p(B) = {\rm e}^{-B^2/2}/\sqrt{2\pi}$, which reduces to $-B'p(B)$ (the result of \citealt{ps74}) when $X\gg1$.

For a wide variety of smoothing filters, power-spectra and barrier shapes, $f_\MS(s)$ remains a good approximation to $ f(s)$ also down to scales on which a substantial fraction of the walks cross with negative slopes \citep{ms12}.  However, it cannot be accurate to arbitrarily small scales since, for a constant barrier, the integral of $f_\MS(s)$ over all $s$ diverges.  This is, of course, a consequence of the fact that multiple upcrossings of the barrier may become important as $s$ increases; they need to be accounted for with the techniques formally described by \cite{ms13a}, which are however difficult to evaluate exactly, or with the excellent and efficient numerical approximation of \cite{ms13b}. However, roughly speaking one may expect these corrections -- from walks with two or more turns -- to be of the order of the square of those introduced by the square bracket term in \eqn{sfs}, which mostly accounts for walks with just one turn. Since these are no larger than $10-15 \%$ over most of the range of interest in cosmology, then $f_\MS$ is accurate up to $1-2\%$ on the small mass side of this range (and exact for large masses), so we will continue with this simpler case.  

\subsection{Non-Gaussian walks}
Motivated by \eqn{dcq} we now consider the problem of finding the first crossing distribution of a barrier of constant height $\delta_c$ by the non-Gaussian variate 
\begin{equation}
  \label{deltaNG}
  \delta\equiv \delta_\L - \beta q_n \,,
  \qquad  \mathrm{with}\ 
  q_n^2 \equiv \sum_{i=1}^n \frac{g_i^2}{n}\,,
\end{equation}
where $\beta\equiv\delta_c/q_c$, and $\delta_\L$ and the $g_i$ are zero-mean Gaussian variates with 
\begin{equation}
  \avg{\!g_i^2\!} = \avg{\!\delta_\L^2\!}\equiv s_\L 
  \qquad {\rm and}\qquad 
  \avg{\!g_i g_j\!} = 0\,;
\end{equation}
the mean and second moment of $\delta$ are thus
\begin{equation}
 \avg{\!\delta\!} = -\beta\,\avg{\!q_n\!} \qquad {\rm and}\qquad
 s\equiv \avg{\!\delta^2\!} = s_\L\,(1 + \beta^2)\,.
\end{equation}
Note that the variance of $\delta$ is $\sigma^2\equiv s - \avg{\!\delta\!}^2$, so this differs from the usual definition of $s$ as the variance of a zero-mean variate. We found it convenient to work in terms of $s$ rather than $\sigma$, as its expression is simpler. Of course, covariance guarantees that the two choices are equivalent, with $f(\sigma^2)=f(s)\,\dd s/\dd \sigma^2$. We will comment further on this point later. 

One can then compute
\begin{equation}
  \Delta = \frac{\delta}{\sqrt{s}}
  =\frac{\Delta_\L-\beta\, Q_n}{\sqrt{1+\beta^2}}\,,
\label{Delta}
\end{equation}
where $\Delta_\L\equiv\delta_\L/\sqrt{s_\L}$ and $Q_n\equiv q_n/\sqrt{s_\L}$. While $\Delta_\L$ is a unit variance Gaussian process, $Q_n$ is (by definition) a chi-variate with $n$ degrees of freedom, whose distribution is
\begin{equation}
  p_{\chi_n}(Q_n) = \frac{2}{Q_n}
  \bigg(\frac{nQ_n^2}{2}\bigg)^{\!\!n/2}
  \frac{\mathrm{e}^{-nQ_n^2/2}}{\Gamma(n/2)}.
\end{equation}
Here $\Gamma$ (not to be confused with the parameter of the walks!) denotes the Gamma function.  The mean of $Q_n$ is 
\begin{equation}
  \avg{\!Q_n\!} = \sqrt{\frac{2}{n}}\,\frac{\Gamma(n/2+1/2)}{\Gamma(n/2)}\,.
\end{equation}

Being the convolution of a Gaussian with a $\chi_n$ variate, $\Delta$ is manifestly a non-Gaussian variate with distribution    
\begin{align}
 \label{pdfNG}
  p(\Delta) &= \int_0^\infty\!\! \dd Q_n\,
  \frac{\mathrm{e}^{-\left(\sqrt{1+\beta^2}\Delta + \beta Q_n\right)^2/2}}{\sqrt{2\pi /(1+\beta^2)}} 
  \,p_{\chi_n}(Q_n) \,.
\end{align}
Although the integral can be evaluated exactly, for $\beta<1$ and $n>1$ it is very well-approximated by a Gaussian with the same mean $\avg{\!\Delta\!}$ and variance $1 - \avg{\!\Delta\!}^2$; this will be useful in the next section.  In addition, it is worth noting that $\avg{\!q_n\!}\propto\sqrt{s_\L}$, so that the variance of $\delta$ is linearly proportional to $s_\L$.

The correlation structure of the $\Delta$ walks is inherited from those of $\Delta_\L$ and $Q_n$.  Since $\Delta_\L$ is Gaussian, its correlation structure is simple (the joint distribution of $\Delta_\L$ and $\Delta_\L'$ factorizes as in equation~\ref{pgauss}), so the issue is the correlation structure of $Q_n$, which is determined by that of $n$ independent Gaussian walks.  
Differentiating $Q_n$ one gets
\begin{equation}
  Q_n' \equiv \frac{\dd Q_n}{\dd s_\L}
   = \sum_{i=1}^n\frac{G_iG_i'}{nQ_n}
\end{equation}
in terms of the unit variance Gaussian variates $G_i\equiv g_i/\sqrt{s_\L}$. Since $\avg{\!G_i'G_j'\!}=\delta_{ij}\avg{\Delta_\L'^2}$ and $\avg{\!G_i'G_j\!}=\avg{\!G_i'Q_n\!}=0$, it follows that 
\begin{equation}
 \avg{\!Q_n'\!}=0 \qquad {\rm and}\qquad 
 \avg{\!Q_n'^2\!} = \avg{\!\Delta_\L'^2\!}/n,
\end{equation}
or equivalently that
\begin{equation}
  \Gamma_q^2 = n\,\Gamma_\L^2\,.
  \label{gammaq}
\end{equation}

To compute the first crossing distribution, one needs the conditional distribution of $Q_n'$ given $Q_n$. The relations above imply that $\avg{Q_n'^{2m}|Q_n}=(2m-1)!!\,\avg{\!Q_n'^2\!}^m$ at fixed $Q_n$ is actually independent of $Q_n$, and $\avg{Q_n'^{2m+1}|Q_n}=0$. Therefore, not only are $Q_n'$ and $Q_n$ uncorrelated (as always), but they are also independent:  their joint probability distribution factorizes, just like in the Gaussian case (equation~\ref{pgauss}). Furthermore, $Q_n'$ is Gaussian, even though $Q_n$ is not. Although this is a new and interesting result in its own right, in the present context it is just a step towards the quantity of real interest.

Now that we know the correlation structure of the non-Gaussian variate $Q_n$, we can address the problem of the $\Delta$ walks. Recalling that $s \equiv \avg{\delta^2} = s_\L\,(1+\beta^2)$, which makes $\dd s/\dd s_\L = s/s_\L$, one has
\begin{equation}
  \Delta'\equiv\frac{\dd\Delta}{\dd s}
  =\bigg(\frac{s_\L}{s}\bigg)^{\!3/2}
  (\DL'-\beta Q_n').
\label{NGvel}
\end{equation}
This shows that $\Delta'$ is proportional to the difference of two zero mean Gaussian variates so it too is Gaussian with mean zero and variance
\begin{equation}
  \avg{\!\Delta'^2\!}=(s_\L/s)^3(1+\beta^2/n)\avg{\!\Delta_\L'^2\!}\,. 
\end{equation}
The relation between the parameter $\Gamma$ for the non-Gaussian walks and the corresponding $\Gamma_\L$ is therefore
\begin{equation}
  \Gamma^2 = \frac{1+\beta^2}{1+\beta^2/n}\, \Gamma_\L^2\,.
 \label{gammaNG}
\end{equation}

Furthermore, since the stochastic variables $\DL'$ and $Q_n'$ are independent of $\DL$ and $Q_n$, so is $\Delta'$ of $\Delta$. This means that $p(\Delta,\Delta')$ factorizes like in the Gaussian case, with $p(\Delta')$ being Gaussian even though $p(\Delta)$ is not.  Therefore, $f_\MS(s)$ is given by \eqn{sfs} with $p_\G(B)$ replaced by $p(B)$ from \eqn{pdfNG}, and $\Gamma_\L$ replaced by $\Gamma$ of \eqn{gammaNG}.  It is remarkable that the structure of the first crossing solution for these non-Gaussian walks is so similar to the Gaussian case.  In particular, for these walks, neglecting the Hermite polynomial terms in equation~(23) of \cite{ms13a} leads to the exact result.

Note that had we chosen to work with $B_\sigma\equiv b/\sigma$ rather than $B$ (i.e. normalizing by the square root of the variance of $\delta$ instead of $\sqrt{s}$), then we would have defined $\Delta_\sigma'\equiv \dd(\delta/\sigma)/\dd\sigma^2$, and hence $\Gamma_\sigma^{-1} \equiv 2\sigma^2\,\sqrt{\avg{\!{\Delta_\sigma'}^2\!}}$.  Although $\Gamma_\sigma\ne \Gamma$,  $f_\MS(s)$ depends not on $\Gamma_\sigma$ but on $X_\sigma\equiv -2\Gamma_\sigma\sigma^2\,\dd B_\sigma/\dd\sigma^2$.  A little algebra shows that 
 $X_\sigma = (\dd B_\sigma/\dd\sigma^2)/\sqrt{\avg{\!{\Delta_\sigma}'^2\!}}
      = (\dd B/\dd s)/\sqrt{\avg{\!\Delta'^2\!}} = X$, so the final answer for $f_\MS(s)$ does not depend on the normalization convention, up to the overall factor $\dd s/\dd\sigma^2$ needed to preserve the covariance of the distribution.

\begin{figure}
 \centering
 \includegraphics[width=0.9\hsize]{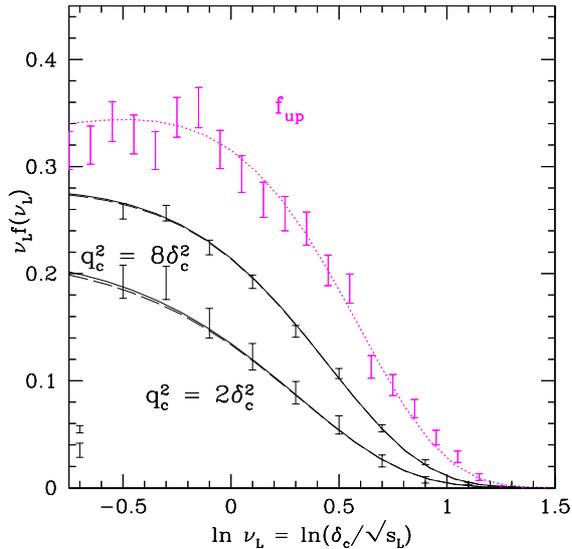}
 \caption{First crossing distribution of a barrier of constant height $\delta_c$ by Gaussian (upper) and non-Gaussian walks (lower) having correlated steps due to tophat smoothing of a CDM $P(k)$.  Results for two values of the parameter $q_c/\delta_c$, which determines the strength of the non-Gaussian component, are shown.   The dotted curve shows \eqn{sfs} for a Gaussian distribution and $\Gamma_\L$, and the solid curves show \eqn{sfs} with the appropriate non-Gaussian distribution (equation~\ref{pdfNG}) and $\Gamma$ (equation~\ref{gammaNG}).  Dashed curves, almost indistinguishable from solid ones, show the result of setting $\Gamma = \Gamma_\L$ when defining $X$ in \eqn{sfs}.}
 \label{chi2fig}
\end{figure}

Finally, we stress that in order to derive this result we have assumed that $\beta$ is constant. Had $\beta$ depended on $s$, then additional terms would appear in \eqn{NGvel} so $\Delta'$ would no longer be independent of $\Delta$.  Nevertheless, expressing all occurrences of $Q_n$ in terms of $\Delta$ using \eqn{Delta}, we could still have said that $p(\Delta'|\Delta)$ is Gaussian, although with a mean value that depends on $\Delta$ and a modified variance.  On the other hand, a scale dependent $\beta(s)$ would likely be signalling that there are hidden variables in the problem that must be made explicit (much like integrating over $q_n$ in the present case would induce a non-vanishing scale dependent mean value for $\delta_\L$, as done by \citealt{st02}).  This is why we have not investigated scale dependent $\beta$ further.

\subsection{Comparison with previous work}
The structure of our solution makes it easy to see why previous approximations to $f(s)$, based on heuristics, worked rather well.  For instance, equation~(A1) of \citet{scs13} is motivated by equation~(13) of \citet{ms12}.  The integral on the right hand side of their equation~(A1) is the same as our \eqn{pdfNG}.  Therefore, their expression for $f(s)$ is the same as ours for $f_\MS(s)$, except that they ignore the difference between $\Gamma_\L$ and $\Gamma$.  This difference is small in the $\beta\ll 1$ limit where they were working, so they found good agreement with their Monte Carlo solutions of this problem.  This approximation is intuitively simple:  the full $f_\MS$ is a sum over the upcrossing distributions for walks crossing a constant barrier of height $\delta_c + \beta\,q_n$ with the contribution associated with $q_n$ being weighted by $p_{\chi_n}(q_n)$.

A slightly different approximation, which yields additional insight, is that of \cite{st02}.  They argued that requiring $\delta \ge \delta_c$ is the same as requiring that $\delta - \beta\,\epsilon_q \ge \delta_c + \beta\, \avg{\!q_n\!}$, where $\epsilon_q \equiv q_n - \avg{\!q_n\!}$.  At large $n$, $p(\epsilon_q)$ becomes approximately Gaussian with mean zero and variance $\avg{\!\epsilon_q^2\!} = s_\L - \avg{\!q_n\!}^2\to s_\L/2n$ as $n\to\infty$.  As a result $\delta - \beta\,\epsilon_q$ is approximately a Gaussian variate with mean zero and variance $\sigma^2 = s_\L\,(1 + \beta^2) - \beta^2\avg{\!q_n\!}^2 \to s_\L\,(1 + \beta^2/2n)$.  
Since $\avg{\!q_n\!}\propto \sqrt{s_\L}$, the quantity on the right hand side is like a deterministic `moving' barrier $b_{\rm eff}(s_\L)$, which the (approximately) Gaussian walks must cross.  Therefore, $f_\MS(s)$ should be well-approximated by \eqn{sfs} with $B = b_{\rm eff}(s_\L)/\sqrt{\sigma^2(s_\L)}$, $X = -2\Gamma_\L \sigma^2\,\dd B/\dd \sigma^2$ and $\Gamma_\L$ equal to the value for the purely Gaussian $\delta_\L$ walks. For example, when $n=5$ then $\avg{\!q_5\!}= (8/3)\sqrt{2s_\L/5\pi}$, so $\sigma^2 = s_\L\,[1 + \beta^2 - \beta^2 (128/45\pi)]$.

This line of reasoning led to equation~(31) of \citet{scs13}, who showed that it was indeed in reasonable agreement with their Monte Carlos.  Like the previous approximation, this one ignores the fact that $\Gamma_\sigma\ne\Gamma_\L$, but in the limit where $\beta^2\ll 1$ this difference matters little.  Strictly speaking, equation~(31) of \citealt{scs13} also ignores the fact that $\avg{\!q_5\!}/\sqrt{s_\L}$ is slightly different from unity, and that $s\ne s_\L$, since $\beta\ll 1$ anyway.  Using the correct mean value removes most of the difference between the solid and dashed curves in their Fig.A1.  And correctly using $\sigma^2$ rather than $s_\L$ makes this differ from their A1 only because it uses the Gaussian approximation instead of the full $p(B)$ -- but we know this is a very good approximation.  

The symbols in Figure~\ref{chi2fig} show the first crossing distribution, obtained by Monte Carlo methods of walks whose steps are correlated because of TopHat smoothing of a CDM power spectrum, for which $\Gamma_\L\approx 1/3$.  We show results for $q_c/\delta_c = 2, 8$ and $\infty$ (the limit in which the shear does not matter, so the non-Gaussian component does not contribute).  The curves in Figure~\ref{chi2fig} show that \eqn{sfs}, with the appropriate value of $\Gamma$, does indeed provide a very good description of the first crossing distribution.  The dashed curves which lie slightly below the solid ones show equation~(A1) of \citet{scs13}.  We argued above that they differ from the solid ones only because they ignore the difference between $\Gamma$ and $\Gamma_\L$:  evidently, this only makes a small difference.  

Stochasticity in the halo formation has also been considered, under the name of stochastic (or diffusing) barriers, by \citet{mr10b} and \citet{ca11}.  In their work, the barrier is assumed to undergo Gaussian uncorrelated fluctuations around a constant (or weakly $s$-dependent) mean, although there is no compelling reason for any of these restrictions (but see \citealt{arsc13} for why they may not be as unrealistic as one might imagine).  The analysis in \cite{cs13} shows how to remove the restrictions on barrier shape as well as to allow for correlated steps.  In effect, our analysis here allows one to also account for physically motivated non-Gaussian scatter around the mean.  However, to justify the statement that our approach is closer to the physics we must first account for the fact that the subset of walks around which collapse occurred is a biased subset of the full ensemble.  \cite{ms12} argued that if the stochasticity around the mean $q$ can be ignored, then this can be accomplished by appropriately weighting each $\delta_\L$ walk \citep[also see][]{ps12}.  Determining the weights when $q$ is also stochastic is the subject of work in progress.


\section{Discussion}\label{discuss}

We showed that the correlation structure of $\chi_n$ walks is remarkably similar to that of Gaussian walks:  the probability of walk height and slope factorize  similarly to how they do for the Gaussian case (equation~\ref{pgauss}).  Therefore, when expressed in terms of the mean-square walk height $s$, the upcrossing distribution $f_\MS(s)$ for $\chi_n$ walks is the same as that for Gaussian walks (equation~\ref{sfs}), but with $p_G\to p_{\chi_n}$ and $\Gamma_\L\to n\,\Gamma_\L$ (equation~\ref{gammaq}).  

We then used this result to derive an exact expression for the upcrossing distribution for walks which are obtained by convolving a Gaussian and a $\chi_n$ variate (equation~\ref{deltaNG}); we argued that this is the problem which is relevant to stochastic barrier models such as \eqn{dcq}, in which the (traceless) shear plays an important role in halo formation.  As for the $\chi_n$ walks, our expression boils down to replacing the Gaussian probability distribution with the non-Gaussian one (equation~\ref{pdfNG}), and rescaling the parameter $\Gamma$ which describes the strength of the correlations between steps (equation~\ref{gammaNG}), in the expression for Gaussian walks (equation~\ref{sfs}).  

The structure of our solution made it easy to see why previous approximations worked as well as they did, and their limitations (Figure~\ref{chi2fig} and related discussion).  In contrast to what previous work assumed, our analysis shows that $\Gamma$ is not the same as the corresponding $\Gamma_\L$ for Gaussian walks.  However, the difference happens to be small for the stochastic barrier models of most interest, since these tend to have the shear (which contributes the non-Gaussian component) being less important than the density, so the effect on $f_\MS$ is small (Figure~\ref{chi2fig}).  

However, even when the non-Gaussian component is large, our expression for $f_\MS$ is much simpler than the corresponding expression for generic non-Gaussian walks \cite[equations~23 or~25 in][]{ms13a}.  Since it yields an excellent approximation to the first crossing distribution (Figure~\ref{chi2fig}) we expect it will find use in analytic models of the cosmic web.  In particular, \cite{sams06} have argued that whereas the `moving' barrier associated with halo formation scales approximately as $\delta_c + 0.5\,\sqrt{s_\L}$, that for sheets scales as $\delta_c - 0.5\,\sqrt{s_\L}$.  They argued that, as a result, the most massive halos are expected to be a substantial fraction of the mass of the filaments they populate, and these filaments a substantial fraction of the sheets in which they are embedded.  Thus, massive halos will appear to be accreting most of their mass from filaments, whereas the growth of lower mass halos may be less obviously anisotropic.  

Recently, upon noting that the $\sqrt{s_\L}$ dependence is primarily due to the shear $q$ (recall that $\avg{q}\propto \sqrt{s_\L}$), \cite{scs13} have shown that this anisotropic growth manifests as nonlocal stochastic bias in the spatial distribution of massive halos.  Since the $\delta_c - 0.5\,\sqrt{s_\L}$ scaling for sheets in the \cite{sams06} model is also primarily due to a dependence on $q$, our analysis here allows one to extend theirs to allow for walks with correlated steps, thus enabling more realistic models of the abundance and clustering of sheets as well as halos.  In particular, \cite{mps12} argued that the Lagrangian bias of halos can be estimated by cross-correlating halos with Hermite-polynomial weighted matter fields.  To isolate the nonlocal effects which come from $q$ one would use the orthogonal (Laguerre) polynomials associated with a $\chi^2_5$ distribution.  This weighted counts estimate should also return the nonlocal bias factors for filaments and sheets.


\section*{Acknowledgments}


MM is supported by the ESA Belgian Federal PRODEX Grant no.~4000103071 and the Wallonia-Brussels Federation grant ARC no.~11/15-040.  He is grateful to the University of Pennsylvania for its hospitality in December 2013.  RKS was supported in part by NASA NNX11A125G.  He is grateful to the Perimeter Institute for its hospitality in September 2013, G. Dollinar and D. Bujatti for their hospitality in Trieste during October 2013, and the Institut Henri Poincare for its hospitality in November 2013.


\label{lastpage}


\bibliography{mybib}{}

\end{document}